\documentclass[12pt,a4paper]{article}

\usepackage[margin=1in]{geometry}
\usepackage[T1]{fontenc}
\usepackage[utf8]{inputenc}
\usepackage{lmodern}
\usepackage{setspace}
\usepackage{graphicx}
\usepackage{booktabs}
\usepackage{longtable}
\usepackage{array}
\usepackage{float}
\usepackage{caption}
\usepackage{natbib}
\usepackage{hyperref}
\usepackage{url}
\usepackage{microtype}

\graphicspath{{./}}
\setcitestyle{authoryear,round}
\hypersetup{
  colorlinks=true,
  linkcolor=black,
  citecolor=black,
  urlcolor=blue,
  pdftitle={Push-Pull Determinants Among Bangladeshi Students Enrolled in NCR Private Universities: A Single-Destination Exploratory Study},
  pdfauthor={Md Millat Hosen and Md Nazmus Sakib}
}

\newcolumntype{L}[1]{>{\raggedright\arraybackslash}p{#1}}
\newcolumntype{C}[1]{>{\centering\arraybackslash}p{#1}}

\title{Push-Pull Determinants Among Bangladeshi Students Enrolled in NCR Private Universities:\\A Single-Destination Exploratory Study}
\author{%
\begin{tabular}{@{}C{0.46\textwidth} C{0.46\textwidth}@{}}
\textbf{Md Millat Hosen} & \textbf{Md Nazmus Sakib}\\
\small Department of Computer Science and Engineering, Sharda University, Greater Noida, Uttar Pradesh, India &
\small Department of Computer Science and Engineering, Sharda University, Greater Noida, Uttar Pradesh, India\\
\small \href{mailto:millat6575@gmail.com}{millat6575@gmail.com} &
\small \href{mailto:2022806985.md@ug.sharda.ac.in}{2022806985.md@ug.sharda.ac.in}\\
\small \href{mailto:2022811342.md@ug.sharda.ac.in}{2022811342.md@ug.sharda.ac.in} &
\end{tabular}}
\date{}

\begin{document}

\maketitle

\begin{abstract}
\noindent
International student mobility from Bangladesh is a significant feature of South Asian higher education, yet India's National Capital Region (NCR) remains an underexplored destination for outbound Bangladeshi students. This exploratory single-destination study examines push-pull factors among Bangladeshi students already enrolled at private universities in India's NCR. A structured online survey was administered to students at Sharda University, Noida International University, and Galgotias University ($n = 63$ total; $n = 56$ retained after quality filtering). Descriptive statistics, K-means clustering, and binary logistic regression were applied within Lee's push-pull theoretical framework \citep{Lee1966}. Preliminary evidence suggests that political and administrative disruptions in Bangladesh's academic calendar were the leading push factor ($M = 3.73$), while geographical and cultural proximity was the strongest pull factor ($M = 3.80$), closely followed by visa accessibility ($M = 3.73$); these two leading pull items did not differ significantly in mean salience. Australia and Germany were tied as the most frequently considered alternative destinations (33.9\% each). Advisory networks influenced 73.2\% of respondents under a broad threshold and 66.1\% under a stricter threshold, primarily for university and course selection. Satisfaction and recommendation intent were positively associated (Pearson $r = .497$; Spearman $\rho = .535$), and infrastructure satisfaction showed the strongest association with high recommendation intent in a five-predictor logistic model (OR = 2.54, 95\% CI [1.09, 5.89]). K-means clustering, validated by one-way ANOVA across six key variables (all $p < .01$), produced three exploratory decision-profile clusters: Comprehensively Motivated, Proximity-Led Enrollers, and Low-Salience Enrollers. The findings suggest that geographic nearness, cultural familiarity, and visa accessibility may operate as a composite accessibility advantage in this short-haul intra-regional corridor.

\vspace{0.5em}
\end{abstract}

\section{Introduction}

Global tertiary-level student mobility is now tracked by international datasets covering major sending and receiving countries \citep{IIEProjectAtlas}. While major Western destinations remain prominent in the internationalization literature, intra-regional mobility, including movement within and from South Asia, is also relevant to understanding contemporary student flows \citep{DeWitAltbach2021}.

Bangladesh represents a compelling case within this shift. UNESCO-based estimates reported by \citet{Rahman2026} place Bangladesh's internationally mobile tertiary student population at approximately 76,000 by 2023, while \citet{Hossain2025} link Bangladeshi students' migration intentions to research opportunities, social expectations, and political instability. India, geographically adjacent and culturally proximate, is one destination within this outflow. Private universities in the Delhi-NCR corridor, including Sharda University, Noida International University (NIU), and Galgotias University, have visible Bangladeshi student communities; based on the lead author's institutional vantage point, Sharda University alone was estimated to enroll more than 250 active Bangladeshi students at the time of this study (institutional estimate, not independently verified against official enrollment records).

Despite this demographic reality, the decision-making processes governing Bangladesh-to-India student mobility are not well characterized in the academic literature. Existing push-pull research is dominated by Western-destination flows \citep{MazzarolSoutar2002,AltbachKnight2007}, while India as a destination for South Asian intra-regional students has received limited systematic attention. Three dimensions are particularly underexplored: (1) the relative salience of push versus pull factors for this specific origin-destination dyad; (2) the role of educational advisory agencies as information intermediaries; and (3) post-enrollment satisfaction dynamics and their relationship to recommendation behavior. The paper's theoretical contribution is to test whether short-haul intra-regional mobility is organized less by discrete, rank-ordered pull factors than by a composite accessibility advantage in which geographic nearness, cultural familiarity, and visa feasibility reinforce one another.

This exploratory paper addresses these gaps with four research questions:
\begin{itemize}
  \item \textbf{RQ1:} Which push factors in Bangladesh most significantly contribute to outward student migration to India?
  \item \textbf{RQ2:} Which pull factors in India most significantly attract Bangladeshi students to the NCR region?
  \item \textbf{RQ3:} To what extent do educational advisory networks influence the university selection of Bangladeshi students?
  \item \textbf{RQ4:} What is the relationship between post-enrollment satisfaction with academic infrastructure and willingness to recommend India as a study destination?
\end{itemize}

The paper proceeds as follows: Section 2 reviews the relevant literature across seven thematic areas; Section 3 describes the study methodology; Section 4 presents results; Section 5 discusses findings in relation to existing theory; Section 6 identifies limitations; and Section 7 concludes with implications and directions for future research.

\section{Literature Review}

\subsection{Push-Pull Theory in International Student Migration}

Everett Lee's \citeyearpar{Lee1966} \textit{A Theory of Migration}, published in \textit{Demography}, remains the foundational framework for understanding geographic mobility including educational migration. Lee proposed that migration decisions are governed by four interacting factor sets: conditions at the origin (push factors), conditions at the destination (pull factors), intervening obstacles, and personal characteristics. The balance of push and pull must overcome inertia and intervening obstacles, among which visa accessibility and geographic distance figure prominently.

\citet{MazzarolSoutar2002} operationalized Lee's framework for international student contexts in a landmark study spanning Indonesia, Taiwan, China, and India. Their \textit{International Journal of Educational Management} paper identified six country-level pull factors: awareness and knowledge of the host country, course availability, cost, social links, the host country environment, and geographic proximity. They positioned social and economic forces at the origin as constituting the push motivation, while the specific destination is determined by pull factor configuration. This paper adopts their operationalization and extends it to the Bangladesh-India dyad.

More recent applications confirm the framework's continued utility. \citet{NikouKadelGutema2025} applied push-pull analysis alongside post-graduation intentions among international students in Finland, finding that career mobility aspirations interact with destination-level factors in shaping preference formation, a finding relevant to interpreting the present study's alternative-destination data.

\subsection{Determinants of Study Destination Choice}

Destination-choice determinants operate across policy, institutional, and social-information layers: visa regimes and post-study work rights shape macro-level feasibility, while institutional cost, course availability, and peer or family information channels shape the alternatives students perceive as realistic \citep{MazzarolSoutar2002,NikouKadelGutema2025}. \citet{Liu2024} discusses visa policy as a factor in international student mobility. In the present study, visa accessibility is therefore treated as an empirical pull factor to be measured rather than assumed in advance.

Cost operates at the meso level. India's official Study in India platform explicitly positions the country as offering high-quality education at competitive cost \citep{StudyInIndia}, making affordability a plausible institutional pull factor for regional students. Bangladesh-specific evidence on domestic university choice similarly finds that distance from home, university location, tuition cost, scholarship availability, family input, peer influence, and campus visits shape institutional selection \citep{IslamShoron2020}; while domestic choice is not equivalent to international destination choice, it supports treating cost, proximity, and social information as locally salient decision dimensions. Cultural and geographic proximity are also consistent with the broader push-pull literature's emphasis on social links, host-country environment, and distance-related barriers \citep{MazzarolSoutar2002}.

\subsection{Bangladesh as an Origin Country}

Bangladesh's higher education system faces compound structural push pressures that students experience through competition, research opportunity constraints, and political uncertainty. \citet{Hossain2025}, in a quantitative study of 325 Bangladeshi students using PLS-SEM, found that research opportunities and societal expectations are significant migration motivators, and, notably, that political instability increases the importance of job adaptation as a mediating factor in migration intention, a finding directly consistent with the present study's identification of political disruption as the leading push factor.

Qualitative evidence also supports this political-governance dimension. In a focus-group study explicitly grounded in push-pull theory, \citet{Zaman2024} found that Bangladeshi students' migration decisions were shaped by governance-related push pressures including political instability, corruption, discrimination, and restricted freedom of expression.

The present study also treats specialized technology education as a possible push mechanism, measured through respondents' perceptions of emerging-program availability in Bangladesh. \citet{Rahman2026}, in a qualitative comparative study of tertiary students in Bangladesh and Nepal, document that migration aspirations are continuously reshaped through encounters with visa regimes, labor markets, and academic cultures. This supports treating student mobility intentions as dynamic responses to both home-country constraints and destination-country opportunity structures.

\subsection{India as a Study Destination}

India's role as a destination for South Asian intra-regional students is shaped by both policy ambition and regional accessibility. The National Education Policy 2020 identifies internationalization as a higher education priority and links it to institutional collaboration, student mobility, and the global visibility of Indian universities \citep{GovernmentIndia2020}. The official Study in India platform similarly frames India as a cost-competitive destination for international students \citep{StudyInIndia}. These policy signals do not by themselves establish student experience or destination quality, but they help explain why Indian private universities increasingly present themselves as viable alternatives to longer-distance Western destinations.

In this paper, Greater Noida institutions including Sharda University, Galgotias University, and Noida International University are treated as active destinations for Bangladeshi students rather than as a representative sample of all Indian higher education. The analytical focus is therefore conditional and corridor-specific: India is examined as experienced by Bangladeshi students already enrolled in NCR private universities.

\subsection{Educational Advisory Networks}

Educational consultancies and advisory agencies function as important information intermediaries in international student recruitment. Student-mobility research has long recognized that migration decisions are mediated not only by formal institutions but also by brokers, family networks, peer information, and imagined post-study trajectories \citep{Baas2010,BrooksWaters2011}. Recent peer-reviewed work has examined students' perceptions and experiences of international recruitment agents, confirming that recruitment-agent relationships remain a live topic within higher education mobility research \citep{SalemMofrehPonniah2025}. In the present study, these actors are treated as part of the decision architecture that students encounter before or alongside direct institutional communication. Quantitative measurement of advisory influence specifically on Bangladeshi students' decision dimensions remains largely absent from prior peer-reviewed work; this study provides initial estimates.

\subsection{Student Satisfaction and Recommendation Intent}

The relationship between student satisfaction and loyalty (recommendation intent) is an established topic in international higher education research. \citet{SultanaMomen2017}, for example, examined international student satisfaction and loyalty in Malaysian and Australian higher learning institutions. The present study extends this satisfaction-loyalty framing to the Bangladesh-India sample, where infrastructure satisfaction and recommendation intent are directly and significantly correlated (Section 5.3).

\subsection{Analytical Methods in Educational Data Mining}

Educational Data Mining (EDM) has been applied to student behavior analysis and profiling across a range of higher education contexts. Logistic regression is among the most widely used classification methods in EDM, valued for interpretability and suitability for binary dependent variables, though its reliability depends on an adequate events-per-variable (EPV) ratio relative to the number of predictors. K-means clustering applied to aggregated Likert-scale scores enables unsupervised grouping of respondents into behaviorally coherent typologies \citep{MichalopoulouSymeonaki2017}. Cluster validity should be assessed using multiple criteria, including the elbow method, silhouette coefficients, and between-cluster significance testing, rather than any single metric in isolation.

\section{Methodology}

\subsection{Research Design}

This study employs a cross-sectional quantitative survey design grounded in Lee's push-pull framework \citep{Lee1966}. A structured online questionnaire was selected for its appropriateness in capturing self-reported decision antecedents and post-enrollment assessments across a defined student population at a single point in time.

\subsection{Survey Instrument}

The questionnaire was developed and deployed on Google Forms. Five sections were included:

\textbf{Section A (Demographics):} Age range, gender, level of study, prior educational curriculum, enrolled university in India, and primary field of study.

\textbf{Section B (Push Factors):} Four Likert items (1 = Strongly Disagree; 5 = Strongly Agree) measuring home-country conditions that influenced the decision to study abroad: limited seat capacity in top-tier domestic public universities (B1); high competition and low acceptance rates in national entrance examinations (B2); absence of specialized or emerging technology programs such as AI and Data Science (B3); and desire to avoid local political or administrative disruptions to the academic calendar (B4).

\textbf{Section C (Pull Factors):} Four Likert items measuring India/NCR-specific attractors: overall affordability of tuition fees at Indian private universities (C1); cost of living in the NCR region relative to Western destinations (C2); geographical and cultural proximity to Bangladesh (C3); and the high success rate and ease of the student visa process (C4).

\textbf{Section D (Advisory Networks and Decision Pathways):} One 5-point scale item measuring the degree to which regional educational advisory agencies influenced the final university choice (D1); a multiple-select item (maximum 2 choices) identifying the type of advisory support received; a multiple-select item identifying alternative countries that were seriously considered; and a single-select item identifying how advisor credibility was verified.

\textbf{Section E (Post-Enrollment Outcomes):} Two Likert items measuring whether academic infrastructure met pre-admission counseling expectations (E1) and willingness to recommend India as a study destination to home-country peers (E2).

Internal consistency of the push and pull subscales was assessed using Cronbach's alpha. The pull subscale (C1--C4) demonstrated excellent internal consistency ($\alpha = .907$), confirming that the four pull items form a cohesive latent construct. The push subscale (B1--B4) returned $\alpha = .652$, below the conventional threshold for a strong unidimensional scale, which is expected because the four push items intentionally measure structurally distinct mechanisms: seat capacity, competitive pressure, curriculum gap, and political disruption. Push items are therefore analyzed individually rather than as a composite index.

Bangla-language descriptions were added beneath each question to improve comprehension and reduce form abandonment. Translation was checked by both authors, who are bilingual in Bangla and English, but no independent professional back-translation was conducted. An informed consent statement at the top of the form disclosed the academic purpose of the study, confirmed data anonymization, and affirmed that no data would be used for commercial purposes. The field ``Science (Pure/Applied)'' was added to Section A's Primary Field of Study question on April 1, 2026, following feedback from a prospective respondent. No Science students had submitted prior to this update, so no existing responses required recoding; nevertheless, this is treated as a minor instrument amendment and is disclosed as a limitation.

\subsection{Data Collection}

The survey was distributed via two channels between March 18 and May 11, 2026: (1) WhatsApp community groups for Bangladeshi students at Sharda University (232-member group) and peer-forwarded sub-groups at NIU and Galgotias; and (2) BCC email outreach to 65 Bangladeshi students via the Sharda University institutional email system. The lead author's former internship role in the International Relations Division (IRD) of Sharda University facilitated initial access to the student distribution network during the data-collection period; the implications of this former role for sampling are addressed in Section 6. A total of 63 responses were received across the collection period. Respondents who selected ``Other'' for their enrolled university ($n = 6$) self-identified as attending other NCR institutions but did not specify the institution name; these institution identities are therefore unknown to the researchers.

\subsection{Data Quality and Cleaning}

Each response was manually reviewed and cross-validated against three exclusion criteria using an automated Python screening procedure:
\begin{enumerate}
  \item \textbf{Straight-lining:} Identical Likert scores across all ten rating items (B1--B4, C1--C4, E1--E2) indicated inattentive response patterns ($n = 6$ excluded).
  \item \textbf{Advisory contradiction:} Simultaneous selection of ``advisory influence $\geq 4$'' and ``None, I applied independently'' constituted a logical inconsistency ($n = 0$ excluded; no responses matched this pattern in the final dataset).
  \item \textbf{Extreme bifurcation:} Responses where all push factors = 5 and all pull factors = 1, or vice versa, were flagged as suspected non-genuine entries ($n = 1$ excluded).
\end{enumerate}

Flagged responses ($n = 7$, 11.1\%) were relocated to a separate sheet (Flagged\_Data.csv) to preserve traceability; they were not deleted. The final analytical dataset comprised \textbf{$n = 56$ clean responses}.

\subsection{Analytical Framework}

\textbf{Descriptive statistics} (means, standard deviations, frequency distributions, cross-tabulations) were computed for all survey sections using Python 3.11 with pandas, numpy, and scipy.stats.

\textbf{K-means clustering} ($k = 3$) was applied to the eight push-pull Likert scores (B1--B4, C1--C4). The optimal $k$ was evaluated using both the elbow method on Within-Cluster Sum of Squares (WCSS) and silhouette coefficient analysis. The silhouette coefficient was highest at $k = 2$ (0.276) and marginally lower at $k = 3$ (0.229) and $k = 4$ (0.242). Stability was also stronger for $k = 2$ (mean adjusted Rand index [ARI] = 1.000 across 20 seeds) than for $k = 3$ (mean ARI = .857, median = .849, minimum = .593). The $k = 3$ solution was nevertheless selected on theoretical usefulness rather than numerical optimality: the three-cluster partition separates Proximity-Led Enrollers from Low-Salience Enrollers, whereas the two-cluster partition collapses them into a broad moderate-salience group. Cluster validity was confirmed by one-way ANOVA on six key variables across clusters (all $p < .01$; see Section 4.7). Analysis was performed using scikit-learn's \texttt{KMeans}.

\textbf{Binary logistic regression} was used to predict high recommendation intent (E2 $\geq 4 = 1$; E2 $\leq 3 = 0$). The threshold of 4 was selected because it represents explicit endorsement on the five-point scale (Agree or Strongly Agree) and produced a near-median split in the outcome (29 events, 51.8\%). To address the limited events-per-variable ratio (EPV = 29/10 = 2.9 in a full 10-predictor model), the primary reported model uses five theoretically motivated predictors: infrastructure satisfaction (E1), cultural proximity (C3), visa accessibility (C4), advisory influence (D1), and political disruption (B4), yielding EPV = 29/5 = 5.8. Although this remains below the conventional EPV $\geq 10$ threshold, the five-predictor model's 5-fold cross-validated accuracy (71.2\% $\pm$ 9.3\%) is virtually identical to the full ten-predictor model (71.4\% $\pm$ 7.0\%), with only marginal reduction in McFadden $R^2$ (.277 vs. .320). Coefficient uncertainty was estimated using an unpenalized standardized logistic model; L2-regularized estimates were used as a performance sensitivity check. All models are framed as exploratory given the modest sample size.

\section{Results}

\subsection{Sample Characteristics}

The analytical sample ($n = 56$) was distributed across Sharda University (76.8\%, $n = 43$), NIU (10.7\%, $n = 6$), other unspecified NCR institutions (10.7\%, $n = 6$), and Galgotias University (1.8\%, $n = 1$). Sharda therefore dominates the sample and should be treated as the main institutional context for interpretation. Male respondents constituted 76.8\% ($n = 43$) of the sample; female respondents 23.2\% ($n = 13$). Gender imbalance is acknowledged as a limitation (Section 6). Undergraduates comprised 92.9\% ($n = 52$) of respondents, with postgraduates at 7.1\% ($n = 4$). The age distribution was concentrated in the 20--22 (46.4\%, $n = 26$) and 23--25 (46.4\%, $n = 26$) brackets, with smaller proportions at 17--19 (1.8\%, $n = 1$) and 26+ (5.4\%, $n = 3$). NCTB curriculum background accounted for 76.8\% ($n = 43$) of respondents, followed by Other curricula (14.3\%, $n = 8$) and English Medium/Cambridge-Edexcel (8.9\%, $n = 5$). Fields of study were distributed across Computer Science/IT (28.6\%, $n = 16$), Engineering (26.8\%, $n = 15$), Arts/Humanities (16.1\%, $n = 9$), Medical and Allied Health (12.5\%, $n = 7$), Business and Commerce (12.5\%, $n = 7$), and Science/Pure Applied (3.6\%, $n = 2$).

\subsection{Push Factor Analysis}

\begin{table}[H]
\centering
\caption{Descriptive Statistics for Push Factors ($n = 56$, scale 1--5)}
\label{tab:push}
\resizebox{\textwidth}{!}{%
\begin{tabular}{L{0.12\textwidth} L{0.55\textwidth} C{0.08\textwidth} C{0.08\textwidth} C{0.10\textwidth}}
\toprule
Code & Factor & $M$ & $SD$ & Rank \\
\midrule
B4 & Political and administrative disruptions to academic calendar & 3.73 & 1.27 & \#1 \\
B3 & Absence of specialized technology programs (AI, Data Science) & 3.27 & 1.36 & \#2 \\
B2 & High competition in national entrance examinations & 2.84 & 1.52 & \#3 \\
B1 & Limited seat capacity in public universities & 2.80 & 1.54 & \#4 \\
\bottomrule
\end{tabular}}
\end{table}

Political and administrative disruption (B4) ranked as the strongest push factor ($M = 3.73$), followed by the absence of specialized technology programs (B3, $M = 3.27$). Competitive examination pressure (B2, $M = 2.84$) and seat capacity (B1, $M = 2.80$) ranked third and fourth, both marginally below the scale midpoint (3.00). The pattern indicates that institutional unpredictability, rather than purely competitive dynamics, is the dominant motivator of outward migration in this sample.

\subsection{Pull Factor Analysis}

\begin{table}[H]
\centering
\caption{Descriptive Statistics for Pull Factors ($n = 56$, scale 1--5)}
\label{tab:pull}
\resizebox{\textwidth}{!}{%
\begin{tabular}{L{0.12\textwidth} L{0.55\textwidth} C{0.08\textwidth} C{0.08\textwidth} C{0.10\textwidth}}
\toprule
Code & Factor & $M$ & $SD$ & Rank \\
\midrule
C3 & Geographical and cultural proximity & 3.80 & 1.23 & \#1 \\
C4 & Visa accessibility and approval rate & 3.73 & 1.27 & \#2 \\
C1 & Tuition affordability & 3.45 & 1.36 & \#3 \\
C2 & Cost of living in NCR region & 3.39 & 1.33 & \#4 \\
\bottomrule
\end{tabular}}
\end{table}

Cultural and geographical proximity (C3, $M = 3.80$) and visa accessibility (C4, $M = 3.73$) led the pull factors, separated by only 0.07 scale points. A paired comparison found no statistically significant mean difference between the two items, $t(55) = 0.53$, $p = .598$, with a bootstrapped 95\% CI for the mean difference of $[-0.20, 0.34]$. Tuition affordability (C1, $M = 3.45$) and cost of living (C2, $M = 3.39$) followed. All four pull factors exceeded the scale midpoint. Logistical accessibility and socio-cultural familiarity together constitute a composite accessibility advantage more decisive than financial considerations alone, though cost remains operationally significant.

\subsection{Alternative Destinations Considered}

Among the 56 clean respondents, 66.1\% ($n = 37$) had seriously considered at least one alternative country before enrolling in India. Australia and Germany were tied as the most frequently cited alternatives (33.9\%, $n = 19$ each), followed by the United Kingdom (26.8\%, $n = 15$), Canada (19.6\%, $n = 11$), Malaysia (5.4\%, $n = 3$), and Turkey (1.8\%, $n = 1$). Thirty-three point nine percent of respondents ($n = 19$) indicated India was their only country of consideration.

\subsection{Advisory Network Patterns}

Advisory influence scores (D1) yielded $M = 2.75$ ($SD = 1.31$). Score distribution: 26.8\% scored 1 (no influence), 7.1\% scored 2, 42.9\% scored 3, 10.7\% scored 4, and 12.5\% scored 5. In aggregate, 73.2\% of respondents reported at least some advisory influence (scores 2--5). If score 2 is treated as minimal rather than substantive influence, 66.1\% still reported clear advisory influence (scores 3--5); 23.2\% reported moderate-to-heavy influence (scores 4--5).

University and course selection was the primary advisory service need (48.2\%, $n = 27$), followed by scholarship application assistance (33.9\%, $n = 19$) and visa processing and documentation (32.1\%, $n = 18$). Independent applicants with no advisory support constituted 26.8\% ($n = 15$).

For credibility verification, 41.1\% ($n = 23$) cross-checked official university websites; 21.4\% ($n = 12$) consulted current students or alumni; 19.6\% ($n = 11$) relied on advisor reputation and track record. Government or embassy portal verification was rare (1.8\%, $n = 1$).

\subsection{Satisfaction and Recommendation Intent}

Infrastructure satisfaction (E1) yielded $M = 3.11$ ($SD = 1.33$), marginally above the scale midpoint of 3.00. Recommendation intent (E2) yielded $M = 3.29$ ($SD = 1.25$), also above the midpoint. A paired $t$-test indicated the 0.18-point gap between E1 and E2 was not statistically significant, $t(55) = -1.03$, $p = .307$. The two measures were moderately and significantly correlated ($r = .497$, $p < .001$), indicating that satisfaction and recommendation intent move together rather than diverging.

Because E1 and E2 are ordinal five-point Likert items, Spearman's rank correlation was also computed as a robustness check. The association remained moderate and significant ($\rho = .535$, $p < .001$). The paired $t$-test and correlations address different questions: the former shows that mean satisfaction and mean recommendation intent are not significantly different, while the latter shows that students with higher satisfaction also tend to report higher recommendation intent.

\subsection{K-Means Clustering: Student Decision Profiles}

K-means clustering ($k = 3$) applied to the eight push-pull Likert scores identified three interpretable student clusters. Cluster selection was based on the WCSS elbow (237.1 at $k = 3$) combined with theoretical interpretability. Silhouette coefficients were $k = 2$: .276, $k = 3$: .229, $k = 4$: .242, and stability across 20 seeds also favored $k = 2$ (mean ARI = 1.000) over $k = 3$ (mean ARI = .857). The $k = 3$ solution was therefore not selected as the numerically strongest partition; it was retained because it is theoretically more useful for this research question. The $k = 2$ solution produced a broad high-salience versus moderate-salience split: Cluster 1 ($n = 27$) had high centroids on nearly all items (B1=3.30, B2=3.48, B3=3.85, B4=4.56, C1=4.48, C2=4.37, C3=4.70, C4=4.56), while Cluster 2 ($n = 29$) had moderate centroids across both push and pull dimensions (B1=2.34, B2=2.24, B3=2.72, B4=2.97, C1=2.48, C2=2.48, C3=2.97, C4=2.97). The $k = 3$ solution separates the theoretically important Proximity-Led Enrollers group from the Low-Salience Enrollers group rather than merging them into one moderate-salience profile. Cluster validity was independently confirmed via one-way ANOVA.

\begin{figure}[H]
\centering
\includegraphics[width=0.92\textwidth]{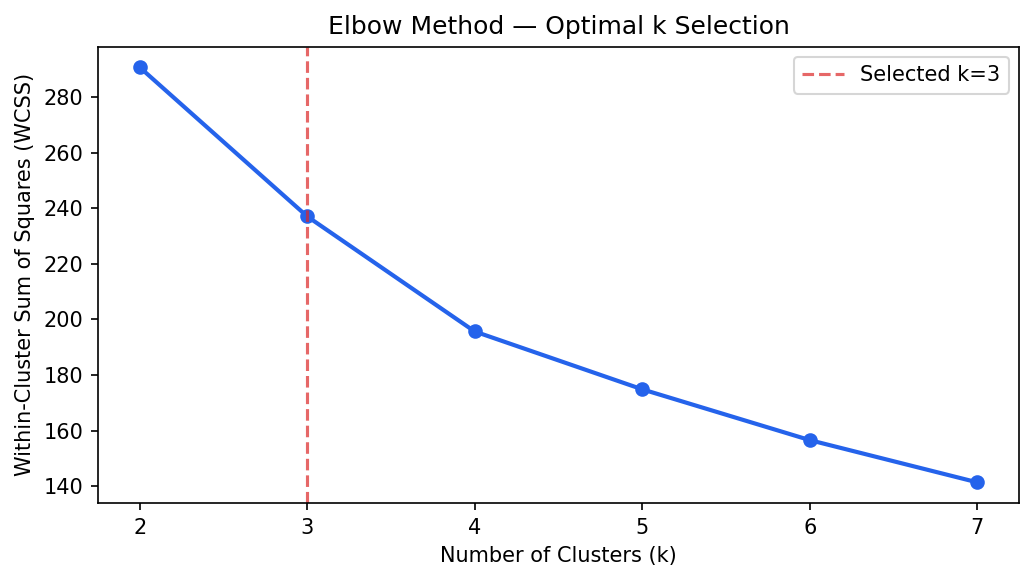}
\caption{Elbow and silhouette diagnostics for K-means cluster selection.}
\label{fig:elbow}
\end{figure}

\begin{table}[H]
\centering
\caption{K-Means Cluster Centroids and ANOVA Validation ($k = 3$, $n = 56$)}
\label{tab:clusters}
\scriptsize
\resizebox{\textwidth}{!}{%
\begin{tabular}{lccc}
\toprule
Feature & Cluster 1 Comprehensively Motivated ($n = 22$, 39.3\%) & Cluster 2 Proximity-Led Enrollers ($n = 19$, 33.9\%) & Cluster 3 Low-Salience Enrollers ($n = 15$, 26.8\%) \\
\midrule
\multicolumn{4}{l}{\textit{Panel A -- Centroids}}\\
B1 Seat capacity & 4.14 & 1.63 & 2.33 \\
B2 Exam competition & 4.23 & 1.47 & 2.53 \\
B3 Tech programs & 3.82 & 3.26 & 2.47 \\
B4 Political disruption & 4.50 & 3.89 & 2.40 \\
C1 Tuition & 4.36 & 3.47 & 2.07 \\
C2 Cost of living & 4.18 & 3.74 & 1.80 \\
C3 Cultural proximity & 4.41 & 4.11 & 2.53 \\
C4 Visa accessibility & 4.41 & 4.21 & 2.13 \\
\midrule
\multicolumn{4}{l}{\textit{Panel B -- Between-Cluster ANOVA}}\\
Variable & \multicolumn{1}{c}{$F(2, 53)$} & \multicolumn{1}{c}{$p$} & \\
C4 Visa accessibility & 38.62 & $< .001$ & \\
B4 Political disruption & 21.73 & $< .001$ & \\
C3 Cultural proximity & 18.45 & $< .001$ & \\
E1 Infrastructure satisfaction & 6.30 & .004 & \\
B3 Tech programs & 5.10 & .009 & \\
E2 Recommendation intent & 5.08 & .010 & \\
\bottomrule
\end{tabular}}
\end{table}

All six between-cluster differences are statistically significant ($p < .01$), confirming that the $k = 3$ cluster structure is empirically valid despite the moderate silhouette coefficient.

\textbf{Cluster 1 -- Comprehensively Motivated (39.3\%):} High scores on every push and pull dimension. B4 (political disruption) peaks at 4.50 within this cluster; these students experienced Bangladesh's institutional constraints comprehensively and found India strongly attractive across all measured pull factors.

\textbf{Cluster 2 -- Proximity-Led Enrollers (33.9\%):} Very low academic competition push (B1 = 1.63, B2 = 1.47) but high pull from cultural proximity (C3 = 4.11) and visa accessibility (C4 = 4.21). These students were not squeezed out by academic competition; they chose India primarily for its positive appeal.

\textbf{Cluster 3 -- Low-Salience Enrollers (26.8\%):} Moderate-to-low scores across all push and pull dimensions. The motivating factors for this cluster likely lie outside the eight measured items, family networks, prior peer presence in India, or advisory recommendations, and constitute a direction for future qualitative inquiry.

\begin{figure}[H]
\centering
\includegraphics[width=0.92\textwidth]{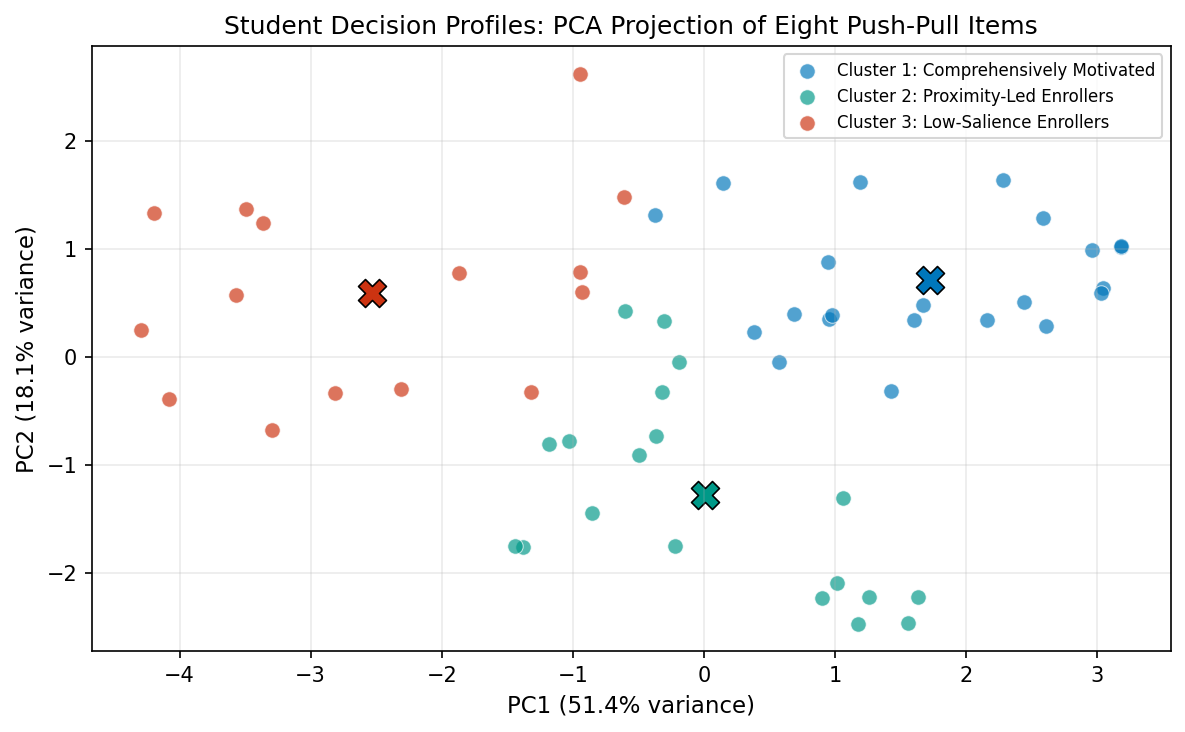}
\caption{PCA projection of the eight push-pull items used for K-means clustering. Points are jittered and semi-transparent to reduce overlap; X markers indicate cluster centroids. PC1 explains 51.4\% of variance and PC2 explains 18.1\%.}
\label{fig:cluster-scatter}
\end{figure}

\subsection{Logistic Regression: Predictors of Recommendation Intent}

\begin{table}[H]
\centering
\caption{Binary Logistic Regression Predicting High Recommendation Intent (E2 $\geq 4$); $n = 56$, events = 29 (51.8\%)}
\label{tab:logistic}
\resizebox{\textwidth}{!}{%
\begin{tabular}{L{0.38\textwidth} C{0.10\textwidth} C{0.10\textwidth} C{0.12\textwidth} C{0.18\textwidth} C{0.08\textwidth}}
\toprule
Predictor & $\beta$ & SE & Odds Ratio & 95\% CI for OR & $p$ \\
\midrule
E1 Infrastructure satisfaction & 0.931 & 0.430 & 2.54 & [1.09, 5.89] & .031 \\
B4 Political disruption & 0.617 & 0.443 & 1.85 & [0.78, 4.41] & .163 \\
C3 Cultural proximity & 0.336 & 0.499 & 1.40 & [0.53, 3.72] & .500 \\
D1 Advisory influence & 0.149 & 0.387 & 1.16 & [0.54, 2.48] & .701 \\
C4 Visa accessibility & 0.057 & 0.489 & 1.06 & [0.41, 2.76] & .907 \\
\bottomrule
\end{tabular}}
\end{table}

\noindent\textit{Primary model: 5 theoretically motivated predictors (EPV = 5.8). Coefficients are standardized, so odds ratios represent the change in odds per one-standard-deviation increase in the predictor. Uncertainty estimates come from an unpenalized standardized logistic model. McFadden pseudo-$R^2$ = .279. The L2-regularized sensitivity model returned McFadden pseudo-$R^2$ = .277, 5-fold CV accuracy = 71.2\% $\pm$ 9.3\%, and training accuracy = 80.4\%.}

The binary outcome contained 29 high recommendation-intent cases (E2 $\geq 4$) and 27 low or neutral cases (E2 $\leq 3$). Infrastructure satisfaction (E1) was the strongest predictor (OR = 2.54, 95\% CI [1.09, 5.89]), indicating that a one-standard-deviation increase in satisfaction substantially increases the odds of high recommendation intent. Political disruption push (B4, OR = 1.85) and cultural proximity pull (C3, OR = 1.40) were positive but imprecisely estimated. Advisory influence (D1) and visa accessibility (C4) showed positive but modest directional associations. Predictor multicollinearity was not severe in the reduced model: variance inflation factors ranged from 1.19 (D1) to 2.50 (C3), with C3--C4 the largest predictor correlation ($r = .68$). Given EPV = 5.8 and the wide confidence intervals, coefficient estimates should be interpreted directionally rather than as precise effect-size measures.

A supplementary 10-predictor model (full push-pull set plus D1 and E1) achieved comparable cross-validated accuracy (71.4\% $\pm$ 7.0\%) and higher apparent McFadden $R^2$ (.320), but with EPV = 2.9 its individual coefficients carry substantial estimation uncertainty and are not reported as the primary result.

\section{Discussion}

\subsection{Cultural Proximity and Visa Accessibility as a Composite Pull Factor}

Cultural and geographical proximity (C3, $M = 3.80$) and visa accessibility (C4, $M = 3.73$) emerged as co-leading pull factors, with no statistically significant mean difference between them ($\Delta = 0.07$, $p = .598$). Their strong correlation ($r = .68$) further suggests that students may experience these items as related dimensions of a shared accessibility construct rather than as independent pull factors. This extends \citet{MazzarolSoutar2002}'s push-pull model in an important direction for short-haul, intra-regional mobility corridors: rather than a single dominant pull factor, the Bangladesh-India dyad is characterized by a \textit{composite accessibility construct}, geographic nearness, cultural familiarity, and logistical ease reinforcing one another. This distinguishes the corridor from longer-distance flows, where cost and institutional reputation typically dominate destination choice.

\subsection{Political Disruption as the Primary Push Factor}

Political and administrative disruptions (B4, $M = 3.73$) ranked as the strongest push factor, ahead of the absence of technology programs (B3, $M = 3.27$). B4's dominance reflects structural unreliability, the inability to depend on continuous, predictable academic calendars, which may be a more urgent enrollment motivator than curricular gaps alone. This interpretation is consistent with prior Bangladesh-focused qualitative work showing that political instability and related governance concerns form part of students' migration reasoning \citep{Zaman2024}. Cluster 1 (Comprehensively Motivated) returned the highest within-cluster B4 centroid (4.50) in the dataset.

Data collection occurred between March and May 2026, approximately 18 months after Bangladesh's major student-led political movement of July-August 2024, which culminated in the resignation of Prime Minister Sheikh Hasina and a significant governmental transition \citep{Ahmed2024}. While the survey instrument measured general perceptions of political disruption rather than that specific event, the elevated B4 scores in the sample may partly reflect its proximity; respondents who enrolled in India in 2024--2025 may have made or consolidated their decision during or immediately after that upheaval. Future longitudinal research should assess whether B4's dominance persists as Bangladesh's political environment stabilizes.

\subsection{Satisfaction Predicts Recommendation: An Aligned, Not Paradoxical, Relationship}

The paired comparison of infrastructure satisfaction (E1, $M = 3.11$) and recommendation intent (E2, $M = 3.29$) yields a gap of 0.18 scale points that does not reach statistical significance ($t(55) = -1.03$, $p = .307$). This mean-comparison result indicates that students' average recommendation intent is not meaningfully higher than their average satisfaction. Separately, the correlation results show co-variation: students with higher satisfaction also tend to report higher recommendation intent, consistent with the satisfaction-drives-loyalty relationship documented by \citet{SultanaMomen2017} in a different institutional context.

The logistic regression confirms this interpretation. Infrastructure satisfaction (E1) is the strongest predictor of high recommendation intent in the five-predictor model (OR = 2.54, 95\% CI [1.09, 5.89]), and the bivariate association is moderate and significant using both Pearson correlation ($r = .497$, $p < .001$) and Spearman rank correlation ($\rho = .535$, $p < .001$). Students who are more satisfied with their academic infrastructure are substantially more likely to recommend India to peers in their home network. The relationship is proportional: satisfaction levels and recommendation levels rise and fall together.

The practical implication is straightforward: for institutions seeking to sustain word-of-mouth enrollment from Bangladesh, improving the accuracy of pre-admission counseling representations, closing the expectation-reality gap that underpins satisfaction scores, is the highest-leverage intervention available. Both satisfaction and recommendation hover near the scale midpoint (3.0), indicating that the institutional experience is adequate but not strongly positive. Improving E1 toward 4.0 would be expected to produce a corresponding increase in E2 according to the logistic model.

\subsection{Advisory Networks as Information Architecture}

Advisory network influence on 73.2\% of students using the broad scores 2--5 threshold, and 66.1\% using the stricter scores 3--5 threshold, positions educational consultancies as a structurally embedded component of the Bangladesh-India enrollment pipeline. University and course selection was the dominant service category (48.2\%). The parallel verification architecture, with 41.1\% of students independently cross-checking official websites and 21.4\% consulting current students, suggests partial trust in advisors rather than wholesale reliance.

\subsection{Alternative Destinations and Competitive Positioning}

Australia and Germany, tied at 33.9\% each, were considered by as many respondents as those who considered India exclusively (also 33.9\%). This positions India not as a first-choice destination competing only with remaining in Bangladesh, but as a frequently chosen alternative to Western destinations, accessible where Western visa and cost constraints are prohibitive. For institutions seeking to improve retention, the relevant competitive frame is therefore ``India vs. eventually transitioning to Australia or Germany,'' not ``India vs. Bangladesh.''

\section{Limitations}

\textbf{Sample size.} The analytical dataset of $n = 56$ supports descriptive and exploratory inferential analysis but remains below the threshold conventionally recommended for stable logistic regression with multiple predictors. The primary regression model was deliberately reduced to five predictors (EPV = 5.8) to partially address this constraint; even so, findings should be interpreted as exploratory and replicated in a larger sample before firm causal claims are made.

\textbf{Cross-validation uncertainty.} Five-fold cross-validation on $n = 56$ leaves approximately 11 observations per held-out fold. The reported 71.2\% $\pm$ 9.3\% cross-validated accuracy should therefore be read as a rough exploratory performance estimate rather than a stable out-of-sample benchmark.

\textbf{Single-destination design and conditional inference.} All 56 respondents enrolled in India; no Bangladeshi students who chose Australia, Germany, or other destinations were surveyed. Consequently, all findings about pull factor salience and satisfaction are conditional on India having been chosen, they cannot be interpreted as unconditional attractors in the full population of Bangladeshi students considering international options. Students who chose Australia likely rate post-study work rights differently; students who chose Germany likely weight tuition cost and language barriers differently. A comparative design sampling Bangladeshi students across multiple destinations would enable stronger inference about India's competitive pull profile. This is the most significant limitation on external validity and constrains the scope of any policy recommendation to India-enrolled students specifically.

\textbf{Researcher-affiliated distribution channel.} The lead author's former internship affiliation with Sharda University's International Relations Division (IRD), which assists Bangladeshi students with enrollment, may have introduced a systematic sampling bias during recruitment and data collection. Students who enrolled through IRD channels may have a different advisory network experience and satisfaction profile than self-directed enrollees or students recruited through other intermediaries. Although the survey was distributed via WhatsApp community groups not restricted to IRD-associated students, this potential confounder cannot be fully ruled out. Future studies should use distribution channels independent of any researcher-affiliated admissions function.

\textbf{Convenience sampling.} Peer-forwarded WhatsApp distribution and institutional email outreach introduce self-selection bias. Respondents who are more engaged in peer networks may not represent the full range of student decision experiences.

\textbf{Gender imbalance.} Female respondents constituted only 23.2\% of the sample. Female students' decision-making patterns may differ systematically from those captured here; conclusions should not be generalized to female Bangladeshi students without targeted replication.

\textbf{Push subscale internal consistency.} The push subscale (B1--B4) returned Cronbach's $\alpha = .652$, reflecting the fact that the four items measure structurally distinct mechanisms rather than a single unidimensional construct. Individual push item means are reported accordingly rather than a composite push score.

\textbf{Ordinal data and K-means assumptions.} K-means clustering treats the Likert scores as approximately interval-level and uses Euclidean distance. This is common in exploratory educational data mining, but it remains an assumption: alternative approaches such as latent class analysis, hierarchical clustering with ordinal distances, or model-based clustering should be tested in larger follow-up samples.

\textbf{Cluster typology and response intensity.} The cluster labels are interpretive rather than causal. Cluster 1's high scores across nearly all measured items may reflect comprehensive motivation, but it may also partly capture acquiescent response style or generally high response intensity. The typology should therefore be treated as a preliminary segmentation to guide future qualitative inquiry, not as a definitive behavioral taxonomy.

\textbf{Retrospective recall bias.} Decision motivations reported post-enrollment may be reconstructed in ways that rationalize the choice made rather than accurately reproducing pre-enrollment reasoning.

\textbf{Mid-survey instrument change and translation validation.} The Science (Pure/Applied) field option was added April 1, 2026, before any Science respondents had submitted, but it remains a protocol amendment. Bangla-language descriptions were checked by the two bilingual authors but were not independently back-translated by a third-party reviewer. Future studies should finalize the instrument and translation validation before data collection begins.

\section{Conclusion}

This exploratory single-destination study characterizes push-pull factors among Bangladeshi students enrolled at private universities in India's NCR region, producing four principal contributions.

First, geographical and cultural proximity together with visa accessibility constitute a composite accessibility advantage that is the dominant pull factor for this origin-destination corridor, more salient than cost or program availability. This refines conventional push-pull interpretation by showing that, in short-haul intra-regional mobility, geographic, cultural, and administrative accessibility may operate as a combined decision structure rather than as separable destination attributes.

Second, political and administrative disruption in Bangladesh is the leading push factor, with its prominence plausibly linked to the 2024 political transition; this should be re-assessed as Bangladesh's political environment evolves.

Third, satisfaction and recommendation intent rise and fall together, with infrastructure satisfaction the strongest predictor of recommendation behavior. Institutions seeking to sustain word-of-mouth enrollment should prioritize closing the gap between pre-admission representations and actual student experience.

Fourth, three statistically validated decision-profile clusters, Comprehensively Motivated, Proximity-Led Enrollers, and Low-Salience Enrollers, provide a typological foundation for differentiated institutional communication and support strategies.

Future research should address the limitations of this exploratory study through larger stratified samples; multi-destination comparative designs sampling Bangladeshi students across India, Australia, and Germany; longitudinal tracking of satisfaction and retention through graduation; gender-stratified sampling; and qualitative follow-up to characterize the motivations of the Low-Salience Enrollers cluster.

\section*{Acknowledgments}

The authors thank the Bangladeshi student communities at Sharda University, Noida International University, and Galgotias University for their voluntary participation. The lead author's former internship role in the International Relations Division of Sharda University facilitated access to student distribution networks during data collection; the implications of this former role for sampling are discussed in Section 6.

\section*{Author Contributions}

\textbf{Md Millat Hosen:} Conceptualization; Methodology; Survey Instrument Design; Data Collection; Formal Analysis; Writing -- Original Draft; Writing -- Review and Editing.

\textbf{Md Nazmus Sakib:} Data Collection; Writing -- Review and Editing.

\section*{Data Availability Statement}

The anonymized survey dataset, flagged-response file, codebook, reproducibility script, requirements file, analysis figures, and manuscript PDF are deposited on Zenodo at \url{https://doi.org/10.5281/zenodo.20760952}. The original Google Forms timestamp column was removed before deposit. The survey instrument is accessible at: \url{https://forms.gle/CFzhqkGCHEUu5y736}

\section*{Ethics Declaration}

This study was conducted as an anonymous, minimal-risk student survey in accordance with the informed consent principles stated in the survey instrument. No personally identifiable information was collected. Participation was entirely voluntary and anonymous, and respondents could stop the survey at any time. No formal institutional ethics committee or IRB approval number was obtained before data collection; this absence of prospective ethics review should be considered a procedural limitation for journal submission. Data were used exclusively for academic research purposes.

\section*{Conflict of Interest Statement}

The authors declare no current conflicts of interest. The lead author no longer holds an internship role in Sharda University's International Relations Division. The lead author's former IRD internship role, disclosed in the Acknowledgments and discussed as a sampling limitation in Section 6, facilitated access to student distribution networks during data collection but did not influence the study's findings, analysis, or conclusions.

\section*{Funding}

This research received no external funding.

\section*{AI Usage Disclosure}

AI tools, including OpenAI ChatGPT/Codex and Anthropic Claude, assisted with literature search structuring, statistical analysis script development, manuscript drafting, LaTeX formatting, and revision checks. The authors reviewed and verified the resulting text, citations, statistical outputs, and interpretations. All research design, data collection decisions, and final interpretation of findings were made by the authors, who take responsibility for the accuracy and integrity of the manuscript.

\newpage
\bibliography{references}

\begin{thebibliography}{19}
\providecommand{\natexlab}[1]{#1}
\providecommand{\url}[1]{\texttt{#1}}
\expandafter\ifx\csname urlstyle\endcsname\relax
  \providecommand{\doi}[1]{doi: #1}\else
  \providecommand{\doi}{doi: \begingroup \urlstyle{rm}\Url}\fi

\bibitem[Ahmed(2024)]{Ahmed2024}
Kaamil Ahmed.
\newblock Why has {Bangladesh}'s prime minister {Sheikh Hasina} resigned and fled?, 2024.
\newblock URL \url{https://www.theguardian.com/world/article/2024/aug/05/why-bangladesh-prime-minister-sheikh-hasina-resigned}.
\newblock The Guardian, August 5.

\bibitem[Altbach and Knight(2007)]{AltbachKnight2007}
Philip~G. Altbach and Jane Knight.
\newblock The internationalization of higher education: Motivations and realities.
\newblock \emph{Journal of Studies in International Education}, 11\penalty0 (3--4):\penalty0 290--305, 2007.
\newblock \doi{10.1177/1028315307303542}.

\bibitem[Baas(2010)]{Baas2010}
Michiel Baas.
\newblock \emph{Imagined Mobility: Migration and Transnationalism among Indian Students in Australia}.
\newblock Anthem Press, London, 2010.

\bibitem[Brooks and Waters(2011)]{BrooksWaters2011}
Rachel Brooks and Johanna Waters.
\newblock \emph{Student Mobilities, Migration and the Internationalization of Higher Education}.
\newblock Palgrave Macmillan, Basingstoke, 2011.
\newblock \doi{10.1057/9780230305588}.

\bibitem[De~Wit and Altbach(2021)]{DeWitAltbach2021}
Hans De~Wit and Philip~G. Altbach.
\newblock Internationalization in higher education: Global trends and recommendations for its future.
\newblock \emph{Policy Reviews in Higher Education}, 5\penalty0 (1):\penalty0 28--46, 2021.
\newblock \doi{10.1080/23322969.2020.1820898}.

\bibitem[{Government of India, Ministry of Education}(2020)]{GovernmentIndia2020}
{Government of India, Ministry of Education}.
\newblock National education policy 2020.
\newblock Technical report, Ministry of Education, Government of India, 2020.
\newblock URL \url{https://www.education.gov.in/sites/upload_files/mhrd/files/NEP_Final_English_0.pdf}.

\bibitem[Hossain et~al.(2025)Hossain, Hasan, Uddin, Yousuf, and Bhuiyan]{Hossain2025}
Rashed Hossain, Md.~Hasibul Hasan, Sayim Uddin, Sharjil~Bin Yousuf, and Mohammad Rakibul~Islam Bhuiyan.
\newblock Determinants of international students' migration intentions for higher education abroad.
\newblock \emph{International Journal of Innovative Research and Scientific Studies}, 8\penalty0 (2):\penalty0 4065--4077, 2025.
\newblock \doi{10.53894/ijirss.v8i2.6231}.
\newblock URL \url{https://www.ijirss.com/index.php/ijirss/article/view/6231}.

\bibitem[{Institute of International Education}(n.d.)]{IIEProjectAtlas}
{Institute of International Education}.
\newblock {Project Atlas}: Explore global data, n.d.
\newblock URL \url{https://www.iie.org/research-initiatives/project-atlas/explore-global-data/}.
\newblock Retrieved June 19, 2026.

\bibitem[Islam and Shoron(2020)]{IslamShoron2020}
Md.~Aminul Islam and Nehal~Hasnain Shoron.
\newblock Factors influencing students' decision making in selecting university in {Bangladesh}.
\newblock \emph{Advanced Journal of Social Science}, 6\penalty0 (1):\penalty0 17--25, 2020.
\newblock \doi{10.21467/ajss.6.1.17-25}.

\bibitem[Lee(1966)]{Lee1966}
Everett~S. Lee.
\newblock A theory of migration.
\newblock \emph{Demography}, 3\penalty0 (1):\penalty0 47--57, 1966.
\newblock \doi{10.2307/2060063}.

\bibitem[Liu(2024)]{Liu2024}
W.~Liu.
\newblock The impact of visa policy on higher education international students.
\newblock \emph{Journal of Education, Humanities and Social Sciences}, 26:\penalty0 92--99, 2024.
\newblock \doi{10.54097/0bkyjg13}.

\bibitem[Mazzarol and Soutar(2002)]{MazzarolSoutar2002}
Tim Mazzarol and Geoffrey~N. Soutar.
\newblock {``Push-Pull''} factors influencing international student destination choice.
\newblock \emph{International Journal of Educational Management}, 16\penalty0 (2):\penalty0 82--90, 2002.
\newblock \doi{10.1108/09513540210418403}.

\bibitem[Michalopoulou and Symeonaki(2017)]{MichalopoulouSymeonaki2017}
Catherine Michalopoulou and Maria Symeonaki.
\newblock Improving {Likert} scale raw scores interpretability with k-means clustering.
\newblock \emph{Bulletin of Sociological Methodology/Bulletin de M{\'e}thodologie Sociologique}, 135\penalty0 (1):\penalty0 101--109, 2017.
\newblock \doi{10.1177/0759106317710863}.

\bibitem[Nikou et~al.(2025)Nikou, Kadel, and Gutema]{NikouKadelGutema2025}
Shahrokh Nikou, Bibek Kadel, and Dandi~Merga Gutema.
\newblock Study destination preference and post-graduation intentions: A push-pull factor theory perspective.
\newblock \emph{Journal of Applied Research in Higher Education}, 17\penalty0 (7):\penalty0 76--96, 2025.
\newblock \doi{10.1108/JARHE-04-2023-0149}.

\bibitem[Rahman et~al.(2026)Rahman, Koirala, Wohab, and Khan]{Rahman2026}
H.~Rahman, S.~Koirala, A.~Wohab, and N.~N. Khan.
\newblock Migration dreams and lived realities: A comparative analysis among tertiary students in {Bangladesh} and {Nepal}.
\newblock \emph{Frontiers in Human Dynamics}, 8:\penalty0 1820900, 2026.
\newblock \doi{10.3389/fhumd.2026.1820900}.

\bibitem[Salem et~al.(2025)Salem, Mofreh, and Ponniah]{SalemMofrehPonniah2025}
S.~Salem, S.~A.~M. Mofreh, and G.~Ponniah.
\newblock Perceptions and experiences with international student recruitment agents: From the students' perspective.
\newblock \emph{Cogent Business \& Management}, 12\penalty0 (1):\penalty0 2555584, 2025.
\newblock \doi{10.1080/23311975.2025.2555584}.

\bibitem[{Study in India}(n.d.)]{StudyInIndia}
{Study in India}.
\newblock Studying abroad made easy with {Study in India} program, n.d.
\newblock URL \url{https://www.studyinindia.gov.in/}.
\newblock Ministry of Education, Government of India. Retrieved June 19, 2026.

\bibitem[Sultana and Momen(2017)]{SultanaMomen2017}
Sarmin Sultana and Abdul Momen.
\newblock International student satisfaction and loyalty: A comparative study of {Malaysian} and {Australian} higher learning institutions.
\newblock \emph{Journal of Intercultural Management}, 9\penalty0 (1):\penalty0 101--142, 2017.
\newblock \doi{10.1515/joim-2017-0005}.

\bibitem[Zaman et~al.(2024)Zaman, Sohel, Obaidullah, Hossen, Rahman, Sifullah, and Sarker]{Zaman2024}
Noshin~Tasnim Zaman, Md.~Salman Sohel, Md. Obaidullah, Md.~Sohrab Hossen, Md.~Toufiqur Rahman, Md.~Khaled Sifullah, and Md. Fouad~Hossain Sarker.
\newblock Factors shaping {Bangladeshi} students' migration decision using push-pull theory: A focus group study.
\newblock \emph{SN Social Sciences}, 4\penalty0 (1):\penalty0 4, 2024.
\newblock \doi{10.1007/s43545-023-00797-2}.

\end{thebibliography}

\end{document}